\documentclass[reprint,aps,prl,superscriptaddress,amsmath,amssymb,floatfix,footinbib,nolongbibliography]{revtex4-1}
\usepackage{graphicx}
\usepackage{xcolor}
\usepackage{bm}
\usepackage{soul}
\usepackage[normalem]{ulem}
\usepackage[colorlinks, linkcolor= blue, citecolor = blue, urlcolor=blue]{hyperref}
\usepackage{comment}
\usepackage{physics}

\usepackage{float}
\usepackage{pifont}
\usepackage{wrapfig}
\usepackage[stable]{footmisc}
\usepackage{array}
\usepackage{gensymb}
\usepackage{multirow}
\usepackage{hhline}
\usepackage{chemformula}
\def\nn{\nonumber}
\def\bea{\begin{eqnarray}}
\def\eea{\end{eqnarray}}
\def\be{\begin{equation}}
\def\ee{\end{equation}}

\def\tc{\textcolor}

\def\bal{\begin{aligned}}
\def\eal{\end{aligned}}

\begin{document}

\title{Symmetry-driven Intrinsic Nonlinear Pure Spin Hall Effect} 
\author{Sayan Sarkar}
\email{sayans21@iitk.ac.in}
\author{Sunit Das}
\email{sunitd@iitk.ac.in}
\author{Amit Agarwal}
\email{amitag@iitk.ac.in}
\affiliation{Department of Physics, Indian Institute of Technology, Kanpur-208016, India.}

\begin{abstract} 
The generation of {\it pure} spin current, spin angular momentum transport without charge flow, is crucial for developing energy-efficient spintronic devices with minimal Joule heating. Here, we introduce the intrinsic nonlinear {\it pure} spin Hall effect (NPSHE), where both linear and second-order charge Hall currents vanish. We show intrinsic second-order spin angular momentum transport in metals and insulators through a detailed analysis of the quantum geometric origin of different spin current contributions. Our comprehensive symmetry analysis identifies 39 magnetic point groups that support NPSHE, providing a foundation for material design and experimental realization. \textcolor{black}{We predict significant nonlinear {\it pure} spin Hall currents in Kramers-Weyl metals even at room temperature, positioning them as potential candidates for NPSHE-based spin-torque devices}. Our work lays a practical pathway for realizing charge-free angular momentum transport for the development of next-generation, energy-efficient spintronic devices.
\end{abstract}

\maketitle
\tc{blue}{\it Introduction---} Harnessing quantum degrees of freedom, such as spin, offers an exciting avenue for designing the next generation of energy-efficient devices and advancing our understanding of fundamental physics. The spin currents, representing the flow of spin angular momentum, enable charge-free information transfer with minimal Joule heating, making them crucial for spintronic applications~\cite{sharma_rmp04, Takahashi_08, Jungwirth_12, Hoffman_13, Sinova_rmp15, saitoh_jap23}. These include spin-torque devices, spin field-effect transistors, and spin Hall nano-oscillators~\cite{Manuel_rpm24, manchon_rmp19, Dutta_apl90, Datta_18, parveen_20, Cheng_prl16}.

However, conventional spin currents are often accompanied by a net charge transport, leading to increased energy dissipation, thermal noise, and electromagnetic interference~\cite{Chien_prl19, Zink, jedema_nature2001, jedema_prb2003}. To address these challenges, {\it pure} spin currents, characterized by charge-less spin transport, have emerged as a foundation for energy-efficient spintronic devices~\cite{Huang_APL2020, Chen_prl19, hoffman_07, lin_18, Althammer_18, gill_24, hirohata_20}. While linear-response mechanisms such as the linear spin Hall effect, spin pumping, and spin Seebeck effects~\cite{perel_pl71, Hirch_prl99, Sinova_prl04, pareek_prl04, Zhang_prl06, Zhang_science06, Sinova_rmp15, Gerrit_prl02, saitoh_apl10} enable {\it pure} spin transport, they are often crystalline symmetry-constrained and can vanish in certain materials. The nonlinear spin Hall effect has been proposed as a symmetry-allowed alternative~\cite{Hamamoto_prb2017, Ghosh_prb2021, Hayami_prb2022, GangSu_prb21, Araki_SR2018, Wang_prl14, GangSu_prb24, Yang_arxiv2024}, yet its quantum geometric origins remain underexplored. More importantly, although the nonlinear {\it pure} spin Hall effect has been explored in specific materials in the optical regime~\cite{Xu_nature2021, Mu_npj2021, dong_24}, a general and comprehensive study in the transport regime is absent, which is critical for nano-scale device integration.

\begin{figure}[t!]
    \centering
    \includegraphics[width=1.0\linewidth]{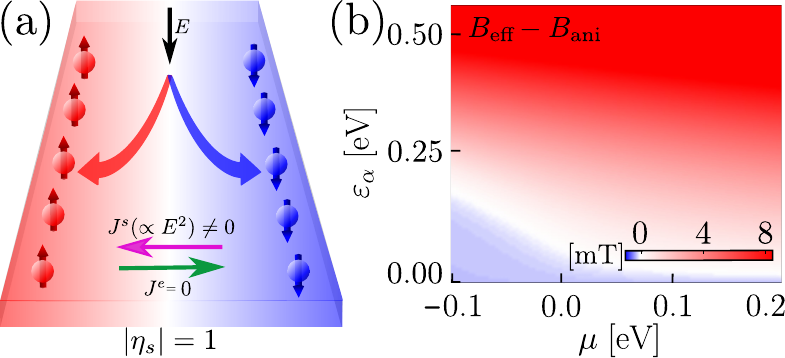}
    \caption{{\bf Schematic of the pure spin Hall effect.} (a) In pure spin Hall effect, equal and opposite spin flows cancel the net charge response, ensuring $|\eta_s| = 1$ [see Eq.~\eqref{purity_param}] and higher energy efficiency. The nonlinear {\it pure} spin Hall effect dominates where the linear response vanishes. (b) The nonlinear spin response generates spin torques manifesting as an effective magnetic field, $B_{\rm eff}$, which can switch magnetization in a ferromagnet when it exceeds the anisotropy field, $B_{\rm ani}$ (see Fig.~\ref{Fig_3} and related discussion). The red region marks the parameter space of the spin-orbit coupling energy scale and chemical potential in Kramers-Weyl metals, where this condition is met. 
    \label{Fig_1}}
\end{figure}

In this Letter, we introduce the nonlinear {\it pure} spin Hall effect (NPSHE), a charge-free spin angular momentum transport mechanism mediated by Bloch electrons [see Fig.~\ref{Fig_1}(a)]. NPSHE dominates in systems \textcolor{black}{along certain} directions where the linear spin response vanishes. Our comprehensive theory of nonlinear spin currents (NSC) reveals key quantum geometric contributions [see Table~\ref{Table_1}] and establishes symmetry-guided design principles to identify materials supporting NPSHE. We predict that $39$ magnetic point groups enable $100\%$ {\it pure} spin transport [$|\eta_s| = 1$, see Eq.~\eqref{purity_param}], in both metallic and insulating systems, with no charge flow up to second-order responses [see Table~\ref{Table_2} in Appendix~B and Table~S3 of SM \cite{Note3}]. We demonstrate significant nonlinear spin Hall currents in Kramers-Weyl metals, which can generate large spin torques, producing an effective magnetic switching field exceeding $10$ mT. As shown in Fig.~\ref{Fig_1}(b), the red region illustrates the parameter space where the NPSHE-generated effective magnetization switching field can suppress the magnetization anisotropy field of a permalloy ferromagnet at room temperature. This positions Kramers-Weyl metals as promising candidates for low-dissipation magnetization switching in spin-transfer-torque devices. Beyond NPSHE, our findings provide a foundation for investigating charge-free angular momentum transport in orbital and magnon currents, paving the way for energy-efficient quantum technologies.


%

\textcolor{blue}{\it Symmetry-guided Design Principles for NPSHE---} We begin by establishing the key principles for systematically identifying materials that support nonlinear spin currents without accompanying charge flow. The second-order NSC density is given by 
\be \label{spin_cur}
J_a^{s_\nu} = \sigma^\nu_{a; bc} E_b E_c, 
\ee
where \(\sigma^\nu_{a; bc}\) is the spin conductivity tensor and $\{\nu,a,b,c\}$ denotes spatial coordinate indices $\{x,y,z\}$. The tensor $\sigma^\nu_{a; bc}$ describes the NSC flowing along $a$, with spins polarized along $\nu$, driven by applied electric fields $E_b$ and $E_c$. To quantify the charge-free nature of spin transport, we define the `{\it purity}' parameter~\cite{gill_24}:
\be \label{purity_param}
\eta_s = \frac{|J^{s_\nu}_a| - |J_a^e|}{|J^{s_\nu}_a| + |J^e_a|}.
\ee 
Here, $J^{s_\nu}_a$ is expressed in units of $\frac{\hbar}{2e}$ and $J^e_a$ is the charge current along the same direction. {\it Pure} spin currents with $J^e_a=0$ correspond to $|\eta_s| = 1$ (see Fig.~\ref{Fig_1}), while impure spin currents satisfy $|\eta_s| < 1$. More significantly, {\it pure} spin currents minimize dissipation, making them energy-efficient. We discuss dissipation in the spin Hall effect and show this in Appendix A. Our Letter focuses on transverse or Hall~\footnote{Note that spin Hall current signifies that the spin current flow direction is perpendicular to the applied electric field direction. Unlike the charge Hall response, the spin Hall response tensor is not guaranteed to be antisymmetric in the spin current and the applied field indices. In general, spin Hall current contains both the symmetric and the antisymmetric components.} spin currents ($a \neq b, c$), as longitudinal spin currents ($a = b$ or $a = c$) are less effective in spin-torque applications due to their alignment with the applied electric field.


Breaking inversion symmetry ($\cal P$) is crucial for finite NSC. However, in $\cal P$-broken systems, both linear and nonlinear charge Hall current can arise, and these must be suppressed to achieve nonlinear {\it pure} spin Hall responses. The linear charge Hall currents can arise from the Drude and the anomalous  Hall effects. The presence of time-reversal symmetry ($\mathcal{T}$) universally prohibits the anomalous Hall effect~\cite{Nagaosa_rmp10}, while the Drude Hall effect can occur in {$\cal T$}-preserving systems~\cite{Tiwari_NL23}. However, it vanishes in systems with rotational symmetry higher than two-fold~(${\cal C}_j^a$ with $j>2$) or mirror symmetry~\cite{Sau_arxiv2023}. Another key advantage of $\mathcal{T}$-symmetric systems is that the NSC is primarily intrinsic, with only extrinsic Drude contribution, as we show later (see Table~\ref{Table_2}). This motivates the search for non-magnetic materials belonging to gray point groups with ${\cal C}_{j > 2}^a$ symmetries, where nonlinear spin Hall responses are permitted, while charge transport is prohibited by symmetry. Furthermore, the presence of $\mathcal{T}$ symmetry suppresses the nonlinear Drude~\cite{Ideue_NP17, Sunit_prb22, Lahiri_prb22} and quantum metric dipole~\cite{Das_Lahiri_2023, Kaplan_prl24} contributions to second-order charge Hall responses. Berry curvature dipole~\cite{Sodemann_prl15} induced nonlinear Hall currents can exist in $\cal T$-symmetric systems. However, they vanish in certain point groups due to the symmetric distribution of Berry curvature over the Fermi surface. This allows us to identify point groups where BCD-induced Hall charge currents are completely suppressed, ensuring pure spin transport. 

Building on these symmetry constraints, we perform a systematic symmetry analysis to identify crystalline materials where NPSHE can be realized experimentally. Our findings identify that $20$ noncentrosymmetric gray point groups and 19 $\cal PT$ symmetric black-and-white point groups can support NPSHE along specific crystallographic directions, with no linear or nonlinear charge currents along the spin current direction. We tabulate the gray point groups and their corresponding NPSHE response tensors in Table~\ref{Table_2} of Appendix B. We also list a few potential candidate materials~\cite{Gallego_MAGNDATA1, Gallego_MAGNDATA2} from each of the gray point groups in Table~\ref{Table_2}. The $\cal PT$-symmetric point groups allowing NPSHE are provided in Sec.~S3 of SM~\cite{Note3}. Depending on the crystalline symmetries and spin current directions, NPSHE in different materials can be collinearly polarized ($\nu=a$), non-collinearly polarized ($\nu \neq a$), or conventional ($\nu \neq a \neq b=c$) spin Hall effect. Some of these point groups may also exhibit intrinsic linear spin Hall effects induced by the spin Berry curvature~\footnote{Note that the Drude component of the linear spin Hall effect is a $\cal T$-odd quantity, thereby vanishes in gray point groups.}, though such responses vanish in two-dimensional gapless systems~\cite{GangSu_prb24}. Further details of symmetry analysis for identifying NPSHE for both non-magnetic and magnetic materials are presented in Sec.~S3 of SM~\cite{Note3}. This provides a symmetry-based foundation for identifying and engineering materials that realize charge-free spin transport. 

\tc{blue}{\it Nonlinear Spin Current---} Building on the symmetry-guided approach, we now develop the theoretical framework for second-order NSC. We calculate the spin current using quantum kinetic theory~\cite{Sekine_prb2017}, where the dynamics of the Bloch electrons are determined by the density matrix $\rho({\bm k},t)$. 
The density matrix $\rho({\bm k},t)$ evolves according to the quantum Liouville equation $i\hbar\partial_t \rho({\bm k},t)=[\mathcal H,\rho({\bm k},t)]$, where $\mathcal{H}=\mathcal{H}_0+\mathcal{H}_E$ is the total Hamiltonian. The unperturbed Hamiltonian $\mathcal{H}_0$ determines the Bloch states $\ket{u_{m\bm{k}}}$ with corresponding eigen energies $\varepsilon_{m\bm{k}}$ as $\mathcal{H}_0 \ket{u_{m\bm{k}}} = \varepsilon_{m\bm{k}} \ket{u_{m\bm{k}}}$. The second term in $\mathcal{H}$ represents the perturbation due to a homogeneous DC electric field, expressed as $\mathcal{H}_E= e \bm{\hat r} \cdot \bm{E}$ with $\hat{\bm r}$ being position operator. To account for relaxation processes, we apply the adiabatic switching-on approximation, assuming ${ E} \equiv { E} e^{\eta t}$ with $\eta \to 1/\tau$~\cite{Sipe_prb00}, where $\tau$ is the relaxation time.

\begin{table}[]
    \begin{center}
    \caption{{\bf Band geometric origin of $\bm k$-resolved spin conductivities.} We express the spin conductivities as ${\tilde \sigma_{a;bc}^\nu} = (e^2/\hbar^2) \sum \rm ( \tau\text{-coeff.} \times \text{Ocu. Fn.} \times  \text{Band Geom.})$, where the summation runs over all relevant band indices. Here, `$\tau\text{-coeff.}$', `Ocu. Fn.' and `Band Geom.' denote the $\tau$-coefficient, occupation function, and band-geometric quantities, respectively. We define $\hbar \omega_{mp}\equiv \varepsilon_{m\bm k}-\varepsilon_{p\bm k}$ and $\hbar v_{mm}^b = \partial_{b} \varepsilon_{m\bm k}$ with $\partial_b \equiv \partial_{k_b}$, while  $f_m^0$ is the equilibrium Fermi distribution function. The covariant derivative is given by  $D_{mp}^b=\partial_b-i(\mathcal{R}_{mm}^b-\mathcal{R}_{pp}^b)$ with $\mathcal{R}_{mp}^a=\bra{u_{m \bm k}}i\partial_{a} \ket{u_{p \bm k}}$ being the Berry connection. Additionally, we define the difference of band-resolved spin current operator and velocities as $\delta j^{\nu a}_{pm}=j^{\nu a}_{pp}-j^{\nu a}_{mm}$ and $\delta v^b_{mp}=v^b_{mm}-v^b_{pp}$, respectively.} 
    \setlength\tabcolsep{0.30 cm}
    \renewcommand{\arraystretch}{2.2}
    \begin{tabular}{c c c c c }
    \hline
    \hline 
    $\tilde{\sigma}_{a;bc}^{\nu}$ & $\tau\text{-coeff.}$ & Ocu. Fn.  &  Band Geom. \\
    \hline
    {$\tilde{\sigma}_{a;bc}^{\nu,\rm SBCP}$} & $\tau^0$   & $\partial_b f_m^0$ & $ 2\, {\rm Im} \dfrac{ j^{\nu a}_{pm}\mathcal{R}_{mp}^c}{\omega_{mp}^2}$\\

    {$\tilde{\sigma}_{a;bc}^{\nu,\rm VI}$} & $\tau^0$ & $ f_m^0 $  & $2 \, {\rm Im}\dfrac{j^{\nu a}_{pm} \delta v^b_{mp} \mathcal{R}_{mp}^c}{\omega_{mp}^3}$  \\

    {$\tilde{\sigma}_{a;bc}^{\nu,\rm SCI}$} & ${\tau^0}$  & ${f_m^0}$ & {${\rm{Re}}\dfrac{\delta j^{\nu a}_{pm}\mathcal{R}^c_{pm}\mathcal{R}^b_{mp}}{\omega_{mp}^2}$} \\

    {$\tilde{\sigma}_{a;bc}^{\nu,\rm Sh}$} & $\tau^0$ & $f_m^0$  & $2 \, {\rm Im}\dfrac{j^{\nu a}_{pm} {D}_{mp}^b \mathcal{R}_{mp}^c}{\omega_{mp}^2}$ \\

    {$\tilde{\sigma}_{a;bc}^{\nu, \rm MB}$} & $\tau^0$ & $f_n^0-f_p^0$  & $-2 \, {\rm Re} \dfrac{j^{\nu a}_{pm} \mathcal{R}_{mn}^b \mathcal{R}_{np}^c}{\omega_{mp}\omega_{np}}$\\

    {$\tilde{\sigma}_{a;bc}^{\nu,\rm SBCD}$} & $\tau^1$ & $\partial_b f_m^0$  &  $2 \, {\rm Re} \dfrac{j^{\nu a}_{pm} \mathcal{R}_{mp}^c}{\omega_{mp}}$\\

    $\tilde{\sigma}_{a;bc}^{\nu,\rm D}$ & $\tau^2$ & $ \partial_b\partial_c f_m^0$  & $j^{\nu a}_{mm}/2$ \\

    \hline 
    \hline
    \end{tabular}
    \label{Table_1}
    \end{center}
\end{table}

The second-order NSC in the steady state is evaluated using the equation, $J^{s_\nu}_{a}=\int_{\bm{k}} {\rm Tr}[\hat{j}^{\nu a} \rho^{(2)}({\bm k},t)]$. Here, $\hat{\rho}^{(2)}({\bm k},t)$ is the second-order (in $E$) density matrix that can be calculated using the standard perturbation technique~\cite{Sekine_prb2017, Sipe_prb00, Debottam_prb24, sarkar_24}. For conciseness, we use the notation  $\int_{\bm k}\equiv \int d^dk/(2\pi)^d$ for a $d$-dimensional system. The spin current operator $\hat{j}^{\nu a}$ is defined as the anti-commutator of the spin ($\hat{s}^\nu$) and the velocity ($\hat{v}^a$) operator: $\hat{j}^{\nu a}=\{\hat{s}^{\nu}, \hat{v}^a\}/2$~\cite{Hamamoto_prb2017, Xu_nature2021, Mu_npj2021}. The resulting NSC density can be expressed as Eq.~\eqref{spin_cur} with
\be
\sigma_{a;bc}^{\nu}(\mu)=\int_{\bm k}\tilde{\sigma}_{a;bc}^{\nu}(\mu,\bm k).
\ee
Here, $\tilde{\sigma}_{a;bc}^{\nu}(\mu,\bm k)$ denotes the $\bm k$-resolved second-order spin conductivity. See Sec.~S1 of SM~\footnote{See Supplemental Material for details on (S1) the calculation of spin currents, (S2) discussion on the definition of the spin current operator, (S3) comprehensive symmetry analysis for NPSHE, and (S4) analytical calculation of nonlinear spin currents in Kramers-Weyl metal.} for a detailed derivation of NSC and a discussion on the definition of the spin current operator~\cite{Niu_prl06, Dimi_npj24, misawa_24}.

We identify seven distinct physical mechanisms contributing to the second-order NSC: five intrinsic mechanisms (independent of the scattering time) and two extrinsic mechanisms (scattering time dependent). The total $\bm k$-resolved spin conductivity can be decomposed as, 
\bea \label{sigma_NSC}
\tilde{\sigma}_{a;bc}^{\nu}(\mu,\bm k) & = & \tilde{\sigma}_{a;bc}^{\nu,\rm SBCP} + \tilde{\sigma}_{a;bc}^{\nu,\rm VI} +\tilde{\sigma}_{a;bc}^{\nu,\rm SCI} +\tilde{\sigma}_{a;bc}^{\nu,\rm Sh} 
 + \tilde{\sigma}_{a;bc}^{\nu,\rm MB} \nn \\
 & + &  \tilde{\sigma}_{a;bc}^{\nu,\rm SBCD}+
 \tilde{\sigma}_{a;bc}^{\nu,\rm D} .
\eea
The different contributions correspond to the i) spin Berry curvature polarizability (SBCP), ii) velocity injection (VI), iii) spin current injection (SCI), iv) shift (Sh), v) multiband (MB), vi) spin Berry curvature dipole (SBCD), and vii) Drude (D) contributions. We list the expressions for each conductivity term in Table~\ref{Table_1}. We discuss the band-geometric origin of the distinct physical mechanism generating NSC of Table~\ref{Table_1} in detail in Appendix C.

All the spin conductivity expressions are U(1) gauge invariant, ensuring their direct applicability in {\it ab-initio} calculations. Notably, $\tilde{\sigma}_{a; bc}^{\nu, \rm SBCP}$, $\tilde{\sigma}_{a; bc}^{\nu, \rm SBCD}$, and $\tilde{\sigma}_{a; bc}^{\nu, \rm D}$ arise from Fermi surface effects, making them finite only in metallic systems. In contrast, the other contributions originate from the Fermi sea and occur in both metals and insulators. Replacing $\hat{s}^\nu$ with $-e {\hat \sigma_0}$ (where ${\hat \sigma_0}$ is the identity matrix) in the spin current operator directly yields nonlinear charge conductivity expressions~\cite{Das_Lahiri_2023}. This highlights a fundamental theoretical connection between nonlinear spin and charge transport mechanisms. 


Tables~\ref{Table_1} and \ref{Table_2} (see Appendix B) provide a comprehensive framework for understanding NPSHE by systematically identifying material candidates and quantifying all contributions to NSC. These form the central results of our Letter. Unlike prior studies on NSC~\cite{Hamamoto_prb2017, Xu_nature2021, Ghosh_prb2021, Lihm_prb2022, Hayami_prb2022, Bhalla_IOP2024, GangSu_prb21, Araki_SR2018, Wang_prl14, Hayami_prb24, Yang_arxiv2024, LiYang_prl21, Fei_24}, which focus on specific mechanisms or systems, our findings systematically encompass all contributions to NSC. By integrating both the Fermi sea and Fermi surface effects with intrinsic and extrinsic mechanisms, our study provides a unified quantum geometric framework for understanding and engineering NSC in real materials.


\tc{blue}{\it NPSHE in Kramers-Weyl Metals---} 
Building on the developed theory and the identification of magnetic point groups that support NPSHE, we demonstrate its occurrence in three-dimensional Kramers-Weyl (KW) metals. KW metals, such as CoSi, RhSi, \ch{K_2Sn_2O_3}, etc., belong to the point group $23.1’$ and exhibit strong spin-orbit coupling~\cite{law21_comm_phys,kang_prb15_transport,samokhin_09, Chang_NM18}. Symmetry constraints in these materials prohibit both linear and nonlinear charge Hall currents (as detailed in Appendix D), thereby enabling NPSHE responses. Specifically, the NPSHE response in KW metals is characterized by $\sigma_{x; yy}^x = \sigma_{y; xx}^y = \sigma_{z;xx}^z = \sigma^z_{z;yy}$, resulting in $100\%$ {\it pure} spin angular momentum flow ($|\eta_s| = 1$). Furthermore, the collinearly polarized linear spin Hall conductivities vanish (see Appendix D), making these nonlinear spin currents the dominant collinear spin response. These collinearly polarized spin currents are crucial for efficient magnetization switching in perpendicularly magnetized materials~\cite{Sinova_rmp15, manchon_rmp19, Yang_arxiv2024}.

\begin{figure}[t]
    \centering
    \includegraphics[width=1.0\linewidth]{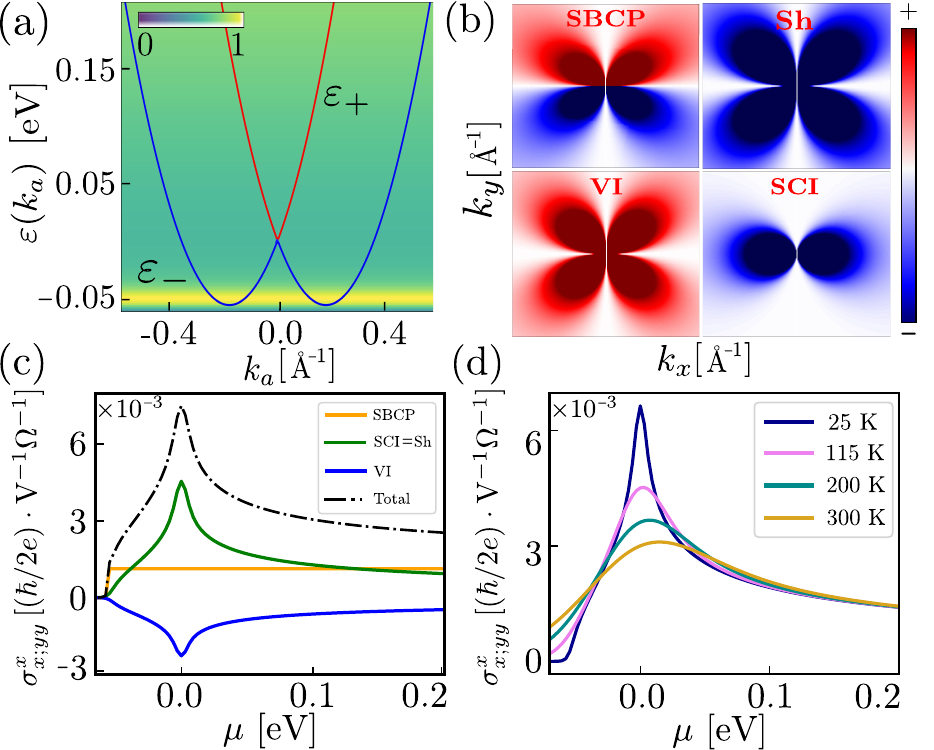}
    \caption{ {\bf Intrinsic NPSHE in Kramers-Weyl metals}. (a) Energy dispersion of a 3D Kramers-Weyl (KW) metal with spin-orbit coupling strength $\alpha=0.7 ~\rm eV \cdot \AA$ and $m^*=1.4 m_e$~\cite{law21_comm_phys} expressed in units of electron mass $m_e$. The background color captures the normalized density of states. (b) Momentum-space distribution of key band geometric quantities (at $k_z=0.005$) -- spin Berry curvature polarizability (SBCP), Shift (Sh), spin current injection (SCI), and velocity injection (VI) -- contributing to $\sigma^x_{x; yy}$. (c) Chemical potential ($\mu$) dependence of different contributions to  $\sigma^x_{x; yy}$ at temperature $T=25~{\rm K}$. The peak at $\mu = 0$ highlights the role of enhanced band geometry near the band crossing points in these responses. (d) Temperature and chemical potential dependence of $\sigma^x_{x; yy}$, demonstrating that KW metals sustain significant NPSHE even at room temperature.}
    \label{Fig_2}
\end{figure}

The low-energy model Hamiltonian for KW metals around the time-reversal invariant momentum is given by~\cite{law21_comm_phys,kang_prb15_transport,samokhin_09, Sunit_prb23, Sougata_prm24}
\be \label{Ham_KW}
{\cal H}=\dfrac{\hbar^2 {\bm k}^2}{2m^*} + \alpha {\bm \sigma}\cdot {\bm k}~.
\ee
Here, $m^*$ is the effective electron mass, $\alpha$ is the spin-orbit coupling parameter, $\bm{ \sigma} = (\sigma_x, \sigma_y, \sigma_z )$ denotes the vector of the Pauli matrices in spin space and ${\bm k} = (k_x,k_y,k_z)$ is the Bloch wave vector. The energy dispersion for the Hamiltonian~\eqref{Ham_KW} is 
$\varepsilon_n = \frac{\hbar^2 k^2}{2 m^*} + n \alpha k $, with $n = \pm $ being the band index. The band dispersion is shown in Fig.~\ref{Fig_2}(a): the inner $n=+$ band (red line) has positive energy, whereas the outer $n=-$ band (blue line) can have both positive and negative energy eigenvalues. The $\bm k$-space distribution of some of the spin band geometric quantities (SBCP, Sh, SCI, and VI) contributing to intrinsic NPSHE are shown in Fig.~\ref{Fig_2}(b). These quantities peak near the Kramers-Weyl point (${\bm k}=0$), highlighting their interband coherence origin, which is enhanced at band touching points.

We analytically calculate the spin conductivities $\sigma_{x;yy}^x$ and $\sigma_{y;xx}^y$ for Hamiltonian~\eqref{Ham_KW}. Various components of $\sigma_{x;yy}^x$, for $\mu>0$ at zero-temperature, are given by 
\begin{subequations}
\bea
&& \sigma^{x,{\rm SBCP}}_{x;yy}  = \frac{\pi e^2 \hbar^2}{15 m^* \alpha^2} ,  \label{sigma_sbcp}\\
&& \sigma^{x,{\rm SCI}}_{x;yy} = \sigma^{x,{\rm Sh}}_{x;yy} = -2\sigma^{x,{\rm VI}}_{x;yy} =\sigma^{x,{\rm SBCP}}_{x;yy} \times {\rm ln}\left(\frac{k_F^-}{k_F^+} \right), ~~~~~\label{sigma_sci_vi_sh}\\
&& \sigma^{x,{\rm D}}_{x;yy} = \frac{ \pi e^2 \tau^2}{15 m^*} [(k_F^-)^2-(k_F^+)^2].  
\eea
\end{subequations}
Here, the Fermi wave vector is given by $k_F^n = -n k_\alpha + \sqrt{k_\alpha^2 + 2 m^* \mu /\hbar^2}$ with $k_\alpha=m^* \alpha/\hbar^2$. Although $\sigma^{x,{\rm SBCP}}_{x;yy}$ appears to diverge as $\alpha \to 0$, this is an artifact of the low-energy model in Eq.~\eqref{Ham_KW}. In reality, when $\alpha \to 0$, the Berry connection vanishes (see Table~\ref{Table_1}), causing $\sigma^{x,{\rm SBCP}}_{x;yy}$ to vanish. The conductivity components of Eq.~\eqref{sigma_sbcp} and \eqref{sigma_sci_vi_sh}, calculated numerically at temperature $T=25$ K, are presented in Fig.~\ref{Fig_2}(c). As a signature of band-geometry-driven phenomena, the NPSHE conductivities rapidly decrease with $\mu$ moving away from the band touching point. This suggests that topological materials with multiple band crossings near the Fermi energy provide a natural platform for optimizing intrinsic NPSHE responses. The Drude contribution to $\sigma^x_{x; yy}$ is presented in Fig.~\ref{Fig_4}(a) of the Appendix D. As a $\cal T$-odd quantity, the SBCD contribution is symmetry-forbidden in KW metals. Furthermore, the MB component is zero due to the two-band model Hamiltonian~\eqref{Ham_KW}. Nonetheless, the MB component can be finite in a realistic tight-binding model of KW metals~\cite{Flicker_prb18, Zahid_prl17, Debasis_prb22} involving multiple bands.

Figure~\ref{Fig_2}(d) shows the variation of $\sigma^x_{x;yy}$ with  $\mu$ for different temperatures. The spin conductivity decreases with increasing temperature, consistent with thermal broadening effects. Remarkably, the strong spin-orbit coupling in KW metals ($\varepsilon_\alpha = m^*\alpha^2 / 2\hbar^2 \approx 45$ meV~\cite{Note3}) allows them to sustain significant spin conductivity even at room temperature. By combining strong spin-orbit coupling with symmetry-protected {\it pure} spin currents, KW metals emerge as potential candidates for next-generation, energy-efficient spin-torque devices operating at room temperature. 


%

\tc{blue}{\it Energy Efficient Magnetization Switching---} In KW metals, the total (including the Drude part from Fig.~4(a) of Appendix-D) nonlinear spin Hall conductivity is estimated to be $\sim 0.05 \, (\hbar/2e) \, \rm V^{-1}\Omega^{-1}$~\cite{Note3}. Under a moderate electric field of $10^6$ $\rm V/m$, the resulting NPSHE current is $J^{s_x}_x \approx 0.5\times10^{11} \, (\hbar/2e) \, \rm A\cdot m^{-2}$. This value is comparable to the linear spin Hall current observed in experiments~\cite{Sinova_rmp15}. Hence, the predicted NPSHE can be readily detected via the inverse spin Hall effect or magneto-optical Kerr spectroscopy. 

\begin{figure}[t]
    \centering
    \includegraphics[width=1.0\linewidth]{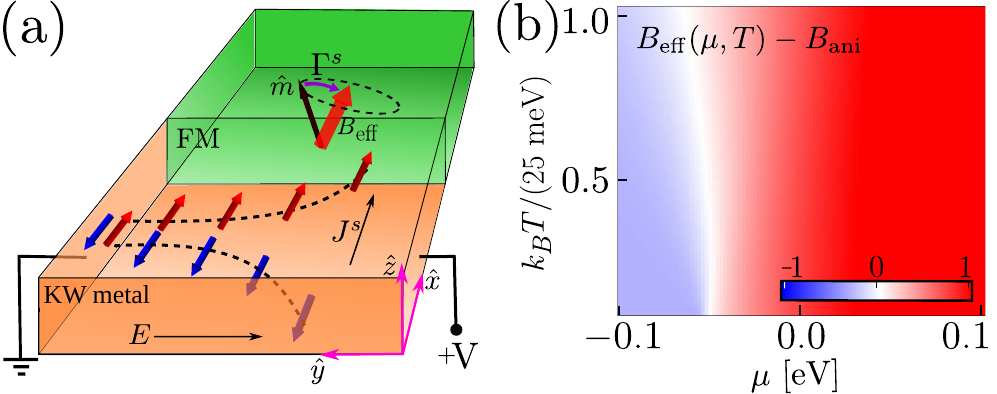}
    \caption{{\bf NPHSE-driven spin torque enables magnetization switching.} (a) Schematic of the spin transfer mechanism in metal-ferromagnet heterojunction. The NPSHE-driven spin current from the Kramers-Weyl metal injects spin angular momentum into the ferromagnetic layer, generating spin torque ($\Gamma^s$) and an effective magnetic field ($B_{\rm eff}$). When $B_{\rm eff}$ exceeds the magnetic anisotropy field ($B_{\rm ani}$), it enables magnetization ($\hat m$) switching in the ferromagnet. (b) Color plot of  $(B_{\rm eff}-B_{\rm ani})$ in the $\mu$--$k_B T$ space, showing the spin transfer torque induced field in permalloy-based ferromagnet. The red region highlights the parameter regime where power-efficient magnetization switching is achieved.}
    \label{Fig_3}
\end{figure}

When injected into a magnetic material, the NPSHE current generates a spin torque per unit magnetization, $|\Gamma^s| = \gamma J^{s_x}_x /(M_s l)$, where $\gamma$ is the gyromagnetic ratio, $M_s$ the saturation magnetization, and $l$ the magnetic layer thickness~\cite{Sinova_rmp15, Liu_prl2011}. This results in an effective field, $B_{\rm eff} = \Gamma^s / \gamma$, acting on localized magnetic moments. Figure~\ref{Fig_3}(a) illustrates this mechanism, where an electrically injected {\it pure} spin current, flowing and polarized along $\hat{x}$, generates a damping-like torque capable of switching the magnetization to align with the $\hat{x}$ direction. Additional details are provided in Sec.~S4 of the SM~\cite{Note3}. The NPSHE-induced spin torque can switch magnetization if $B_{\rm eff}$ surpasses the anisotropy field $B_{\rm ani}$. 

For a permalloy-based ferromagnet (FM), we have $M_s \sim 6.4 \times 10^5~{\rm A/m}$~\cite{Krivorotov_science2005} and $B_{\rm ani}\sim 0.25$ mT~\cite{celinski_apl03, Peng_APL2019, Husain_ACSnano2020, Xie_APL2021, Bera_prb24}, and assume a thickness of $l \sim 5$ nm (see Fig.~\ref{Fig_3} for the schematic of the setup).  As shown in Fig.~\ref{Fig_3}(a), $(B_{\rm eff}-B_{\rm ani})$ varies with $\mu$ and temperature $T$, revealing a red shaded region where $B_{\rm eff} > B_{\rm ani}$, enabling energy-efficient magnetization switching even at room temperature. To further probe room-temperature magnetization switching, Fig.~\ref{Fig_1}(b) presents the variation of $(B_{\rm eff}-B_{\rm ani})$ at $T=300$ K as a function of the spin-orbit coupling energy scale ($\varepsilon_\alpha$) and $\mu$. Remarkably, a wide range of realistic spin-orbit coupling parameters enables magnetization switching, with $B_{\rm eff}$ reaching up to 15 mT. This underscores the potential and utility of NPSHE-driven spin torque in KW metals for realizing room-temperature, energy-efficient magnetization switching in FM. 

\tc{blue}{\it Conclusion---} We have conducted an exhaustive symmetry analysis to identify magnetic point groups and materials supporting NPSHE, enabling 100\% pure spin angular momentum transport. Our comprehensive theory of nonlinear spin currents includes both intrinsic and extrinsic contributions and highlights their quantum geometric origins. We demonstrate a significant NPSHE in Kramers-Weyl metals at room temperature, establishing them as efficient spin torque generators for low-dissipation magnetization switching. This paves the way for next-generation, energy-efficient, charge-free spin Hall devices operating at room temperature~\cite{hirohata_20, Jungwirth_12}.

Beyond spin transport, our work motivates the exploration of analogous charge-free transport mechanisms, including orbital and valley currents~\cite{Culcer_adv24, Kamal_prl24}. Given the similar symmetry properties of spin and orbital angular momentum, our analysis establishes the groundwork for realizing {\it pure} orbital Hall currents, charge-free orbital transfer torques, and their applications in future spin-orbitronic technologies.

\textcolor{blue}{\it Acknowledgments---} We acknowledge many fruitful discussions with Debottam Mandal and Kamal Das. We also thank Koushik Ghorai for help with Figure 1(a). S. S. acknowledges IIT Kanpur for funding support. S. D. is supported by the Prime Minister's Research Fellowship under the Ministry of Education, Government of India.  A. A. acknowledges funding from the Core Research Grant by ANRF (Sanction No. CRG/2023/007003), Department of Science and Technology, India.

\appendix

\textcolor{blue}{\it Appendix A: Dissipation in spin Hall effect---} 
The heat dissipation in the spin Hall effect has three main contributions~\cite{Taniguchi_16}. The first is Joule heating due to applied current or voltage. The second contribution arises from the inverse spin Hall effect contribution to the dissipation. The third contribution is given by the Joule heating due to the charge Hall current along the spin current direction. Combining these and using the spin current `{\it purity}' parameter $\eta_s$, we find that the heat generation rate per unit volume can be expressed as,  
%
\bea 
\frac{\partial Q}{\partial t}  &  =&  \sigma^{\rm D}_{b;b} E_b^2  + 2 \sigma^{\rm D}_{a;b} E_b^2 (J^{s_\nu}_a/J_b^e)^2 \tanh(\frac{l}{2 l_d}) \nn \\
&&  + ~J^{s_\nu}_a \frac{1-\eta_s}{1+\eta_s} E_b~.
\eea 
Here, $J_b^e$ is the applied charge current density, and $E_b$ is the electric field along $b$-direction, whereas the generated spin Hall current is $J^{s_\nu}_a$ with $a\neq b$. $\sigma^{\rm D}_{b;b}$ is the longitudinal charge Drude conductivity. We denoted $l$ as the thickness of the spin current producing material, and $l_d$ is the spin diffusion length. The $\eta_s$  dependent dissipation contribution vanishes for the {\it pure} spin current as $\eta_s \to 1$, leading to relatively more energy-efficient switching dynamics and spin torques.

\textcolor{blue}{\it Appendix B: Detailed Symmetry Analysis---}
In this Appendix, we list all the non-magnetic gray point groups supporting {\it pure} nonlinear spin Hall current, with vanishing linear and nonlinear charge Hall effects. We classify all the allowed tensor components of NPSHE responses and a few candidate materials extracted from the MAGNDATA~\cite{Gallego_MAGNDATA1, Gallego_MAGNDATA2}  in Table~\ref{Table_2}.

\tc{blue}{\it Appendix C: Band Geometric Origins of NSC---} 
In this Appendix, we highlight 
the band-geometric origin of the distinct physical mechanism generating NSC in Eq.~\eqref{sigma_NSC} and Table~\ref{Table_2}. \\
i) {\it Spin Berry Connection Polarizability (SBCP)}-$\tilde{\sigma}_{a; bc}^{\nu,\rm SBCP}$: This Fermi surface contribution arises from electric field induced correction to expectation value of the spin current operator and is proportional to spin Berry connection polarizability defined as $2\, {\rm Im}[j^{\nu a}_{pm}\mathcal{R}_{mp}^c/{\omega_{mp}^2}]$ \cite{Yang_arxiv2024, GangSu_prb24}. \\
ii) {\it Velocity Injection (VI)}-$\tilde{\sigma}_{a; bc}^{\nu,\rm VI}$: It is an intrinsic contribution to the spin conductivity proportional to band velocity difference ($\propto  v^b_{mm} - v^b_{pp}$). This is analogous to the velocity injection-based second-order nonlinear charge currents~\cite{Das_Lahiri_2023, Kamal_prl22, Sipe_prb00, Aversa_prb95}. 

\begin{table}[t]
    \begin{center}
    \caption{{\bf Nonmagnetic crystallographic point groups supporting nonlinear pure spin Hall effect}. This table lists the point groups that allow nonlinear {\it pure} spin Hall currents with no linear or nonlinear charge currents, along with potential material candidates~\cite{Gallego_MAGNDATA1, Gallego_MAGNDATA2}. For this analysis, we have applied the electric field in the $xy$ plane, and the spin conductivities are symmetric in the electric-field indices. We provide a list of magnetic point groups supporting the NPSHE in Sec.~S3 of SM.} %
    \renewcommand{\arraystretch}{1.5}
    \setlength\tabcolsep{0.05cm}
    \begin{tabular}{c|c|c} 
    \hline \hline
    MPGs &  $\sigma^{\nu}_{a;bc}$ Components & Candidate  \\
     & ($\sigma^{\nu}_{a;bc}\equiv\nu abc$) & Materials\\
    \hline \hline
    $222.1',~432.1'$  &  & \ch{Tb3NbO7},~\ch{RhSi}  \\ 
     -$42m.1',~422.1'$ & $xxyy,yyxx,zzxx,zzyy$ & $\alpha$-\ch{Mn},~\ch{Ho2Ge2O7} \\
    -$43m.1',~23.1'$ &  & \ch{NdBiPt},~\ch{CoSi},\\
    \hline
    $mm2.1'$  & $xyxx,yxyy$ & \ch{CePdAl3},\\
    \hline
    $4mm.1',6mm.1'$  & $xyxx,yxyy$ & \ch{NdBPt3,~PrBPt3}\\
    \hline
    $32.1'$  & $xxyy,yyxx,zyxx,xzxy,$ & $\alpha$-\ch{HgS},\\
    & $yzxx,yzyy,zzxx,zzyy$ & \\
    \hline
    $-6m2.1'$   & $zxyy,xzxx,xzyy,yzxy$ & 2H-\ch{MoS2}\\
    \hline
    $622.1'$ & $xxyy,yyxx,zzxx,zzyy$ & \ch{CeSe2},~\ch{CoNb3S6}\\
    \hline
    $6.1'$  & $xxyy,yxyy,xyxx,yyxx,$ & $\ch{Dy3CuGeS7}$\\
    \hline
    $-6.1'$  & $zxyy,zyxx$ & \ch{Nb9As7Pd} \\
    \hline
    $3m.1'$  & $yxyy,xyxx,zxyy,xzxx,$ & \ch{BiTeI},\\
    & $xzyy,yzxy$ & ~\ch{PtBi2} \\
    \hline
    $3.1'$  & $xxyy,yxyy,zxyy,xyxx,$ & $\ch{Ni3TeO6}$ \\
     & $yyxx,zyxx$ & \\
    \hline
    $4.1'$  & $xxyy,xyxx,yxyy,yyxx$ & $\ch{TiPb9O11}$\\
    \hline
    $-4.1'$ & $xxyy,yxyy,xyxx,yyxx,$ & $\ch{Al2CdTe4}$\\
    \hline
    $m.1'$  & $yxyy,xyxx,zyxx,yzyy$ & $\ch{GdBiPt}$\\
    \hline
     $2.1'$  & $xxyy,zxyy,yyxx,xzyy,$ & $\ch{BaNiF4}$,~$\ch{MnWO4}$ \\
     & $zzyy$ &  \\
    \hline
    \hline
    \end{tabular}
    \label{Table_2}
    \end{center}
\end{table}

iii) {\it Spin Current Injection (SCI)}-$\tilde{\sigma}_{a; bc}^{\nu,\rm SCI}$: This contribution stems from the difference of spin current expectation values ($\propto  j^{\nu a}_{pp} - j^{\nu a}_{mm}$) between two bands. Note that $\tilde{\sigma}_{a; bc}^{\nu,\rm SCI}$ is the spin counterpart to the conventional velocity injection contribution for nonlinear charge current.  \\
%
%
iv) {\it Shift Mechanism (Sh)}-$\tilde{\sigma}_{a; bc}^{\nu,\rm Sh}$: The real space shift of the wave-packets, which is represented by the shift vector [$A^{bc}_{mp}=i\partial_b\log{\mathcal{R}_{mp}^{c}}+(\mathcal{R}_{mm}^b-\mathcal{R}_{pp}^b)$]~\cite{Sipe_prb00} where $D_{mp}^b \mathcal{R}_{mp}^c=-i \mathcal{R}_{mp}^c A^{bc}_{mp}$,  gives rise to the $\tilde{\sigma}_{a; bc}^{\nu,\rm Sh}$. Physically, this arises due to different shifts of wave packets associated with opposite spins~\cite{Nagaosa_prb17}. \\
v) {\it Multi-Band (MB)}-$\tilde{\sigma}_{a; bc}^{\nu,\rm MB}$: The multi-band contribution arises from virtual transitions across multiple bands and is finite only in systems with more than two bands. \\
vi) {\it Spin Berry Curvature Dipole (SBCD)}-$\tilde{\sigma}_{a; bc}^{\nu,\rm SBCD}$: Originating from the first moment of the spin Berry curvature at the Fermi surface, this term is proportional to $2\, {\rm Re} [\partial_b (j^{\nu a}_{pm} \mathcal{R}_{mp}^c/{\omega_{mp}})]$. It is analogous to the Berry curvature dipole-induced nonlinear charge response ~\cite{Sodemann_prl15, Sinha_NP22}.  \\
vii) {\it Drude Contribution}-$\tilde{\sigma}_{a; bc}^{\nu,\rm D}$: The electric field induced shift of the Fermi surface generates this extrinsic intraband component of NSC. It is proportional to the quadrupole moment of the intraband spin current operator: $\int_{\bm k} j^{\nu a}_{mm} (\partial_b \partial_c f_m^0) = \int_{\bm k} f_m^0 (\partial_b\partial_c j^{\nu a}_{mm} ) $~\cite{Hamamoto_prb2017}. 

%
\begin{figure}[t]
    \centering
    \includegraphics[width=.9\linewidth]{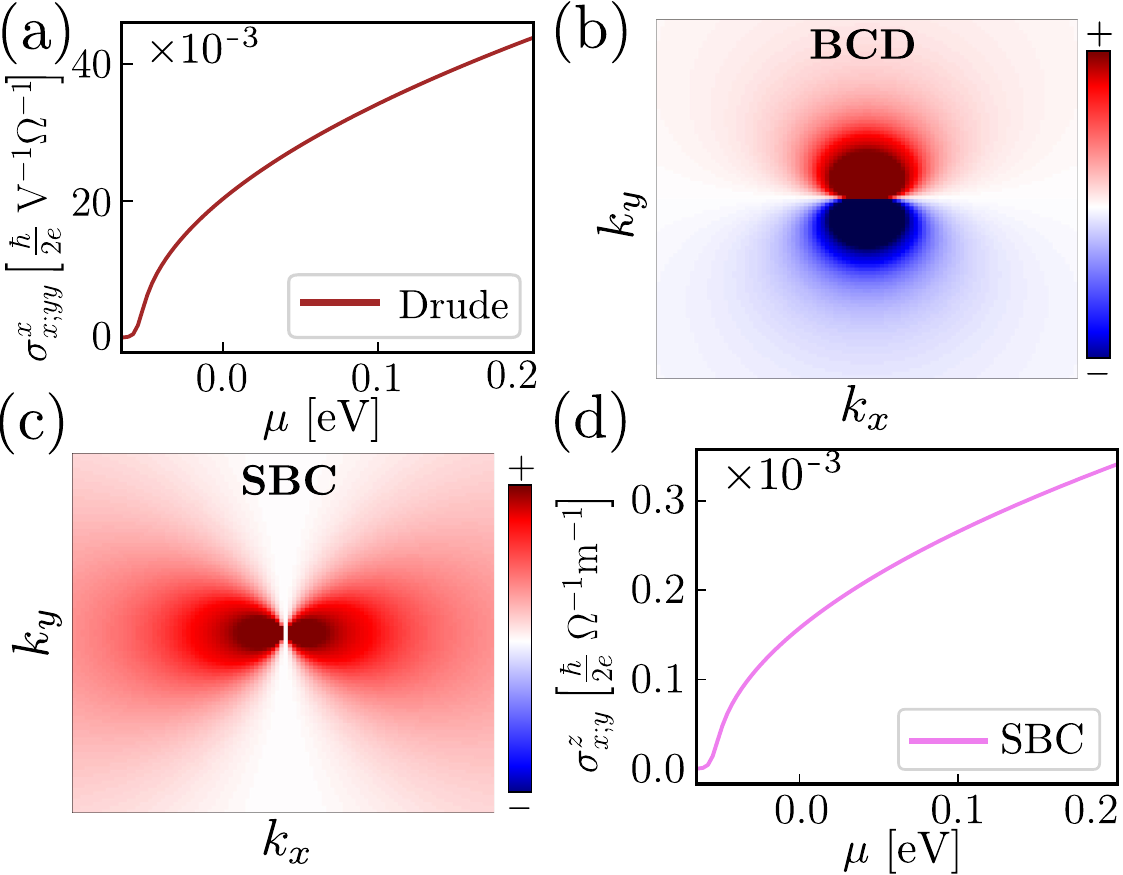}
    \caption{(a) The Drude component of the NPSHE as a function of $\mu$ in KW metals, calculated at $T=25$ K, assuming $\tau= 0.01$ ps. (b) and (c) The Berry curvature dipole density ($\Omega^{z} v^y$) and spin Berry curvature $\Omega^{zxy}$ distribution in $k_x$-$k_y$ space (at $k_z=0.005$) for the $n=+$ band. (d) The spin Berry curvature induced linear spin Hall conductivity $\sigma^z_{x;y}$ as a function of $\mu$.}
    \label{Fig_4}
\end{figure}

\textcolor{blue}{\it Appendix D: No Charge Hall Currents and Linear Spin Transport in KW Metals---} In this Appendix, we show that the collinearly polarized linear spin Hall effects do not occur, while conventional spin Hall responses remain finite in KW metals. Consequently, collinearly polarized nonlinear spin Hall currents dominate as the primary collinear {\it pure} spin Hall response in KW metals.

The eigenstates of Hamiltonian~\eqref{Ham_KW} are given by $\ket{u}_{+}^{T} = [\cos(\theta_k/2), 
e^{i \phi_k} \sin(\theta_k/2)]$ and $
\ket{u}_{-}^T = [\sin(\theta_k/2), -e^{i \phi_k} \cos(\theta_k/2)]$, with  $\cos\theta_k \equiv k_z/k$ and $\tan\phi_k \equiv k_y/k_x$. The velocity along the $a$-direction is $v^a_n = \frac{\hbar k_a}{m^*} +n  \frac{\alpha k_a}{\hbar k}$, and the Berry curvature is ${\bm \Omega}_n =- n \frac{{\bm k}}{2k^3}$. The linear anomalous Hall effect vanishes due to the time-reversal symmetry of the Hamiltonian~\eqref{Ham_KW}. This is because the Berry curvature, ${\bm \Omega}_n$, is an odd function of $\bm{k}$, and its integral over $k$-space vanishes.

The $\cal T$-symmetric linear Drude Hall conductivity,
\be
\sigma_{a;b}^{\rm D} = -\frac{e^2 \tau}{\hbar^2} \sum_{n } \int_{\bm k} v^a_n v^b_n (\partial_{\varepsilon_n} f_n^0),
\ee
also vanishes because $v^x v^y \propto k_x k_y$, $v^x v^z \propto k_x k_y$, and $v^y v^z \propto k_y k_z$ are odd in momentum and integrate to zero. This also follows from the rotational symmetry of KW metals.

Next, we examine the possibility of finite $\cal T$-even nonlinear Hall currents. The $\cal T$-even nonlinear Hall conductivity arising from the Berry curvature dipole (BCD) can be written as~\cite{Sodemann_prl15} 
\be 
\sigma^{\rm BCD}_{a; bc} = -\frac{e^3 \tau}{\hbar}  \epsilon_{abd} \Lambda_{dc}~~\text{with}~~ \Lambda_{dc} = \sum_n \int_{\bm k} \Omega_n^d v^c_n (\partial_{\varepsilon_n} f_n^0).
\ee 
Here, $\epsilon_{abd}$ is the anti-symmetric Levi-Civita tensor. For $\sigma^{\rm BCD}_{x;yy}$, the term $\Lambda_{zy}$ involves the integration of $\Omega_n^z v^y_n$, which is proportional to $k_z k_y$ and thus vanishes upon $k$-space integration. 
Similarly, the BCD current in other directions also vanishes due to the \textcolor{black}{antisymmetric distribution of Berry curvature dipole over the Fermi surfaces, see Fig.~\ref{Fig_4}(b).} 

Given the absence of charge Hall responses in KW metals, we now examine the linear spin Hall conductivity, which has two contributions: the $\cal T$-odd Drude component and the $\cal T$-even spin Berry curvature component, see Sec.~S1 of SM~\cite{Note3}. Due to $\cal T$-symmetry, the former vanishes in KW metals. Now, the $\cal T$-even spin Berry curvature (SBC) contribution to linear SHE can be expressed as 
\be 
\sigma^{\nu}_{a;b} = \frac{e}{\hbar} \sum_{n} \int_{\bm k} \Omega^{\nu a b}_{n}f_n^0,
\ee 
where spin Berry curvature is given by $\Omega^{\nu a b}_n = -2 \, {\rm Im} \sum_{p \neq n}\frac{j^{\nu a}_{np} v_{pn}^b }{(\varepsilon_n-\varepsilon_p)^2}$. For an electric field applied in $y$-direction, the collinearly polarized spin Hall conductivities are given by $\sigma^x_{x;y}$ and $\sigma^z_{z;y}$. The corresponding spin Berry curvature are found to be $\Omega^{zzy}_n= -2\Omega^{xxy}_n = n \dfrac{\hbar k_x k_z}{4m^* \alpha k^3}$. Evidently, these are odd functions of $k_z$ and $k_x$, which vanish upon integration. Similarly, other components such as $\Omega^{yyx}_n$, $\Omega^{zxx}_n$, $\Omega^{zyy}_n$, etc., can be shown to vanish, ensuring the absence of collinearly polarized linear spin Hall responses in KW metals. However, the conventional spin Hall responses where the spin polarization, spin current flow, and the applied electric field are mutually perpendicular to each other can survive. We find such responses are characterized by  $\sigma^z_{x;y}=\sigma^x_{y;z} = \sigma^y_{z;x}$ and $\sigma^z_{y;x} = \sigma^x_{z;y} = \sigma^y_{x;z}$. The spin Berry curvature (SBC) component $\Omega^{zxy}_n$ for the $n$-th band has the $k$-resolved expression as $\Omega^{zxy}_n=n\frac{h^2 k_x^2}{2m^*\alpha k^3}$. It is plotted numerically in the $k_x$-$k_y$ plane in Fig.~\ref{Fig_4}(c). In the zero temperature limit, the $\sigma^z_{x;y}$ is given by
\be
\sigma^z_{x;y} = - \sigma^z_{y;x} = \dfrac{8\pi e k_\alpha}{3\hbar}\sqrt{1+{\mu}/{\varepsilon_\alpha}}.
\ee
The numerically calculated variation of $\sigma^z_{x;y}$ with $\mu$ at $T=25$ K is presented in Fig.~\ref{Fig_4}(d). 

Remarkably, because of vanishing charge Hall responses in KW metals, the linear spin Hall responses also represent {\it pure} spin angular momentum flow. Assuming an electric field of $E\sim 10^6$ V/m, the linear spin Hall current is estimated to be $ \sim 10^2~ (\hbar/{2e})\rm \, A\cdot m^{-2}$. Note that the nonlinear {\it pure} spin Hall current is much larger than the linear spin Hall current. Thus, the NPSHE discussed in the main text dominates the {\it pure} spin transport in Kramers-Weyl metals.

\bibliography{refs}

\end{document}